\documentstyle[prb,twocolumn,aps,epsf,psfig]{revtex}
\begin{document}
\twocolumn[\hsize\textwidth\columnwidth\hsize\csname
@twocolumnfalse\endcsname
\draft
\title{Phase separation frustrated by the long range Coulomb  
interaction II: Applications} 
\author{J. Lorenzana,$^{1,2}$ C. Castellani,$^1$ and C. Di Castro$^1$}
\address{
$^1$ Dipartimento di Fisica, Universit\`a di Roma "La Sapienza'' and  
Istituto Nazionale di Fisica della Materia, 
Unit\`a di Roma I,\\
Piazzale A. Moro 2, I-00185 Roma, Italy.}
\address{
$^2$ Consejo Nacional de 
Investigaciones Cient\'ificas y Tecnicas, Centro At\'omico Bariloche,
8400 S. C. de Bariloche, Argentina }
\date{\today}
\maketitle
\begin{abstract}
The theory of first order density-driven phase transitions
with frustration due to the long range Coulomb  (LRC) interaction
develop on paper I of this series is applied to the following
physical systems: i)  the low density electron gas
ii) electronic phase separation  in the low density three dimensional
$t-J$ model iii) in the  manganites  near the charge ordered phase.
We work in the approximation that the density 
within each phase is uniform  and we assume that the system
separates in spherical drops of one phase hosted by the other phase
with the distance between drops and the drop radius much larger 
than the interparticle distance. 
For i) we study a well known apparent instability related to a
negative compressibility at low densities. We show that this does
 not lead to macroscopic drop formation as one could expect
 naively and the system is stable from this point of view.  
For ii)  we find that the LRC interaction significantly 
modifies the phase diagram favoring uniform phases and mixed states
of antiferromagnetic (AF) regions  surrounded by metallic regions over 
AF regions surrounded by empty space. For iii) we show that 
the dependence of local densities of the phases on the overall density
found in paper I  gives a non-monotonous behavior of the Curie
temperature on doping in agreement with experiments. 
\end{abstract}
\vskip2pc]

\narrowtext
\section{Introduction}
 In the last decades continuous progress in the characterization 
and preparation  of complex compounds have produced a variety 
of systems with very reach phase diagrams when  the concentration 
of some dopant is varied. Notable  examples includes  doped cuprates and
manganites where one finds different phases as the electronic density 
is varied. Quite generally and  
in analogy with  familiar first order phase transitions,
 like the ice-liquid transition, it is   
natural to ask under what conditions one can find ranges of
global electronic density with  phase
separation among the many electronic phases that one finds in these 
materials.\cite{mul92,sig93} 

This problem has arisen  naturally for doped Mott
insulators\cite{low94,eme90c} and Fermi liquid instabilities\cite{cas95b} 
 in the context of the cuprates and also in the related problem of doped magnetic semiconductors.\cite{nag83} 

It is by now settled
that close to the Mott transition there is a
natural tendency for the system to phase separate in insulator and
metallic phases with different densities.
 This tendency is frustrated by the long range 
interaction which tends to favor uniform phases.  As a result the
system can choose to phase separate in a inhomogeneous state 
with islands of one phase in the other phase keeping long scale
neutrality. The same phenomena can occur in a variety of
situations and in particular in the doped magnetoresistant
manganites phase separation at various scales is observed in different
regions of the phase digram.\cite{mor99} Also there is evidence that the 
2-dimensional (2D) electron gas phase separates at low densities.\cite{ila00}

In paper I of this series (hereafter referred as I) we have presented
a theory of phase separation frustrated by the long range Coulomb
(LRC) and in the presence of a surface energy cost. We showed that if
 the Coulomb and surface energy cost are not two strong the phase
 separated state survives but it  is inhomogeneous. 
In certain global density range drops of one phase  ($A$) are formed and
 hosted by the other phase  ($B$).
The free energy per unit volume reads:
\begin{equation}
  \label{eq:fdx}
  f=(1-x)f_A(n_A)+x f_B(n_B)  + e_m
\end{equation}
where $x$ is the volume fraction. The first two terms are 
the bulk  contribution of the $A$ and $B$ phases and the last term is
the mixing energy  
\begin{equation}
  \label{eq:em}
  e_m= \left[\frac{\sigma^2 e^2 (n_B-n_A)^2}{\epsilon_0}\right]^{1/3} u(x)
\end{equation}
 where $u(x)$ is a  geometric factor which in the case  of drops is:
\begin{equation}
  \label{eq:udx}
  u(x)=3^{5/3}\left(\frac{\pi}{10}\right)^{1/3} x
 (2-3x^{1/3}+x)^{1/3}
\end{equation}
The mixing energy  
includes the surface energy cost and the electrostatic cost.  

 In our computations we have assumed that
 the   scales of the inhomogeneities  is much larger than the
 interparticle distance. This study is complementary to others which
 have considered the opposite limit (frustrated phase separation
   at a scale comparable to the interparticle distance) to explain 
phenomena like the  striped states in cuprates.\cite{low94,cas95b}

We have considered spherical drops as done by Nagaev and collaborators in the 
context of doped magnetic semiconductors in general and of 
manganites in particular.\cite{nag83,nag98} However we obtain
similar results in other geometries like a periodic array of layers
and we believe that for any reasonable geometry similar behavior for
thermodynamic quantities  would be obtained.

To illustrate the generality of theses ideas in this paper we consider 
some relevant applications to open problems in condensed matter. 

It is well known that the low density electron gas has a negative 
compressibility.\cite{mah90} We discuss
the fundamental problem of the  stability of the electron gas and of 
the Wigner crystal at low density against a bubble 
phase (Sec.~\ref{sec:wigner}). The system is shown to be stable against 
this kind of phenomena showing explicitly that a negative compressibility
can be observed in this system because the LRC interactions makes it stable. 
Interestingly negative compressibility has been measured in the 2D electron 
gas.\cite{eis92sha96}

 To make a link with the problem of phase
separation in doped Mott insulators we  consider frustrated  
 PS in the $t-J$ model  (Sec.~\ref{sec:stron}). This is one of the 
simplest models used in the context of high temperature superconductors
where frustrated phase separation is believe to play an important role. 
We illustrate the importance of the LRC forces in determining the phase
 diagram.

Finally we study the problem of the phase separation 
in the  Manganites  between a ferromagnetic metallic phase 
and a charge ordered phase (Sec.~\ref{sec:man}). This problem
illustrates nicely how the anomalous behavior of local densities found
in I can reflect in measurable quantities. We propose an explanation for the 
anomalous behavior of the Curie temperature close to a charge ordered
 instability. i.e. the Curie temperature decreases as the 
charge ordered instability is approached. 
We conclude with a summary of the main results (Sec.~\ref{sec:con}).

\section{Stability of the  jellium model  }
\label{sec:wigner}

Here we discuss the case of a system of electrons in a uniform
rigid background usually called the ``jellium'' model. Although we find that
drops do not form in this case, this first discussion is very useful to 
illustrate the range of applicability of the present ideas.

The problem is the following:
It is well known that a low density electron gas has a negative
electronic  compressibility.\cite{mah90} Will this lead to
drop formation?. 

To describe in an approximate way the electronic energy one can 
use the Wigner interpolation formula for the correlation energy.
In this approximation the ground state energy per particle at zero temperature
is given by:\cite{mah90}  
\begin{equation}
  \label{eq:ery}
E_{el}/Ry=\frac{2.2099}{r_s^2}-\frac{0.9163}{r_s} -\frac{0.88}{r_s+7.8}
\end{equation}
where the  first term is the kinetic energy, the second term is the
 exchange energy, and the  last term is the correlation energy. Here 
$r_s= [3/(4 \pi n)]^{1/3}/a_0$ and $a_0$ is Bohr's radius. 

\begin{figure}[tbp]
\epsfxsize=9cm
$$
\epsfbox{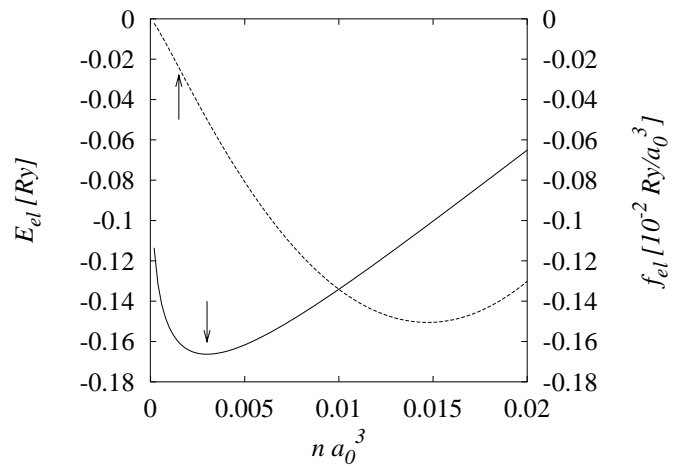}
$$
\caption{Energy per particle (left axis, solid line) and per unit 
volume (right axis, dashed line) 
as a function of density obtained from Wigner interpolation formula.
The vertical arrows indicate the density at which the pressure is
zero (lower one) and the density  at which the jellium model 
contribution to the compressibility becomes zero (upper one).
In the later case the corresponding change of curvature is almost 
indistinguishable to the bare eye. 
}
\label{fig:wigner}
\end{figure}

The energy per unit volume is $f_{el}(n)=E_{el} n$. 
In Fig.~\ref{fig:wigner} we 
plot $f_{el}$ and $E_{el}$ as a function of density. These curves can be 
interpreted in
two different ways. If the background compressibility is given only by the 
electrostatic self energy (already included) then the curves represent the 
total energy of the system, background plus electrons. We call this the 
compressible background case. Two different criteria give 
thermodynamic instability for the  compressible background case.
First for $n a_0^3<0.0015$ (up arrow)  the compressibility 
is negative. More importantly  for  $n a_0^3<0.003\equiv n_{el}^0$
(down arrow) the pressure  is negative. 
The latter means that  if the system is prepared with a density lower than 
$n_{el}^0$, then electrons and background will relax
to a self bound system with a lower volume and 
$n= n_{el}^0$ (from now on we shall measure densities in units of Bohr inverse volume $a_0^3$).
 We can consider this result as due to the usual Maxwell construction
 (MC) argument applied to phase separation between the 
electronic system+background  and the vacuum. In fact it is easy to see that
$n= n_{el}^0$ satisfy  a MC in which the MC line intersects the origin.  
By putting an external pressure densities higher than 
$n_{el}^0$ become physically accessible. 
The  numerical value of $ n_{el}^0$ can change when more accurate 
forms of the correlation energy are consider
but the qualitative picture will remain the same.

The above interpretation is not useful in real situations  where 
forces other than the electrostatic one can constrain the  background
to have a certain density. This leads to a second interpretation of 
 the curves in Fig.~\ref{fig:wigner}. 
Since the background has an additional non-electrostatic contributions
 to the compressibility 
(for example coming from core-core repulsion of the atoms), the 
plots of  Fig.~\ref{fig:wigner}  are not any more 
the total energy of the (neutral) system as function of density. 

A simple hypothesis to describe such a system is to assume in the model 
that the substrate is completely rigid.  
We call this the incompressible background case. 
The curves in  Fig.~\ref{fig:wigner} represent then the energy of many
different realization of this system each with a different
density. The energy including the electrostatic cost
to change the density of the background from one system to the other
but excludes the  (infinite) non-electrostatic energy cost to change
the  density of the background. 

Now the total density is fixed at some value, the 
compressibility  of the whole system is infinite and the above 
 instability criteria  do not apply any more. 
This however does not guarantee  the stability of the system. One can 
imagine that the system may be unstable towards an inhomogeneous phase with
electron rich and electron poor bubbles in the uniform fixed background. 
 
We analyze below  the case in which the  electron 
poor regions have zero electron density. In principle we can work as
in I  
with a quadratic expansion of the free energy around the MC case
however the free energy has now a simple form which can be dealt with 
analytically.

\subsection{Electron drops in  background }

The compressible background case suggests that the system has the tendency to 
separate in electron rich  regions and regions  of zero electron density. 
We will take the $A$ and $B$ phases of I to be the background with no 
electrons and  the background with an undercompensated density
of electrons respectively.
 Consider first the case of low densities for the $B$ phase. 
We can take for the bulk drop  free energy the energy of a classical 
Wigner crystal i.e. 
the leading $1/r_s$ term in Eq.~(\ref{eq:ery}) for large  $r_s$. 
This microscopic Wigner crystal should not be confused with the mesoscopic
Wigner crystal that the drops would form. 
 
It is instructive to write the free energy in the following way: 
\begin{equation}
  \label{eq:dwiginvac}
f=\frac{6\pi e^2n}{5 }   \left\{-  n_{el} r_{el}^2+ n_{el} R_d^2
\left[ 2 - 3 {\left( \frac{n}{n_{el}} \right) }^{1/3} 
+ \frac{n}{n_{el}} \right] \right\}
\end{equation}
The volume fraction of the $B$ phase $x$, has been eliminated by using the 
 constraint in the density given by  $n=x n_{el}$ where 
$n_{el}\equiv3/(4 \pi r_{el}^3)\equiv n_B$ and $n_A=0$.
The first term in the brackets comes from the classical Wigner  crystal
contribution  (the leading term in $f_B\equiv f_{el}$ at low density)  
and the second term is the mixing energy contribution computed in I.
 The latter contains the electrostatic bubble contribution and the surface 
energy contribution. The radius is not a free parameter but
 $R_d\equiv R_d(n,n_{el},\sigma)$ is the drop radius that 
minimizes the free energy.

Notice that the drop of electrons is not neutral since the 
 density of electrons is larger than the compensating 
background i.e. within the drops we are dealing with a charged Wigner crystal 
of electrons in contrast with the usual neutral Wigner crystal of
electrons. On the other hand 
 in the computation of the charging energies of the drops in I
we have assumed  for simplicity that the density 
is uniform inside the drop. 
In the appendix A of paper I we compute the correction to the electrostatic
 energy 
due to the non uniform electronic density at the microscopic scale  
as it should be for a charged Wigner crystal and conclude 
that this only changes numerical  factors, which are not important for the 
present  analysis.

Since the drop radius has already been minimized one is left with the
density inside the drops (or equivalently with $r_{el}$) to be minimized.
$R_d$ depends on density explicitly and indirectly on $\sigma(n_{el})$. 
If the last term in the curly brackets grows with $ n_{el}$ faster than 
$n_{el} r_{el}^2 $, the minimum occurs at $n_{el} \rightarrow n$ i.e. $x=1$.  
This corresponds to the uniform case. If instead 
$n_{el} r_{el}^2 $ grows faster one finds a solution with 
$n_{el}\rightarrow \infty$, the term in the square brackets become a
constant and  clearly $ r_{el}>R_d $. In this case  the mesoscopic bubble 
model is clearly not adequate. In order for  both term to balance exactly 
one finds that the surface energy
has to fulfill the relation $\sigma \sim e^2 n_{el}^{2/3}/r_{el}$. 
If one estimates the surface energy as a characteristic energy density 
($n_{el} e^2/r_{el}$) times a characteristic length 
($r_{el}\sim n_{el}^{-1/3}$) one can conclude on 
dimensional arguments that this surface energy is precisely the one of a 
Wigner crystal. Smaller surface energies give drops which are too small  
for the mesoscopic treatment and larger surface energies 
 give no drops at all.

 What about the other  contributions to the bulk 
free energy in Eq.~(\ref{eq:ery}) which will become important 
as the density inside the drop becomes large? 
They only make the drop bulk term less negative so 
an even smaller value of the drop radius is needed to stabilize the
 drop solution. From this point of view we can conclude that
mesoscopic or macroscopic drops of electron gas are  not possible. 

 The only dubious case could be close to $x=1$  ($n_{el} \rightarrow n$).
Since in this case the term in the square brackets  can be very small.  
 In principle this allows for 
large drops without paying too much mixing energy. 
However in this region Eq.~(\ref{eq:dwiginvac}) is not strictly valid since 
the volume fraction is close to 1 and one  has to revert the geometry
as done below.

\subsection{ Drops of empty background (voids) }
We consider the possibility of formation of  drops of zero electronic 
density (voids) hosted by  electron rich regions with density $n_{el}$. 

We look again to the limit of the classical Wigner crystal. Now
$x$ will represent the fraction of empty electronic volume. The constraint 
in the density is given by $n=(1-x) n_{el}$ and the free energy reads:
\begin{eqnarray}
  \label{eq:empty}
f&=&\frac{6\pi e^2}{5 }   \Bigg\{-  n n_{el} r_{el}^2 \\
&+&
 ( n_{el}-n) n_{el} R_d^2   
\left[ 3 - 3 {\left( 1-\frac{n}{n_{el}} \right)}^{1/3} 
-\frac{n}{n_{el}} \right]   \Bigg\} \nonumber
\end{eqnarray}
We see that if  $n_{el} \rightarrow n$ we can get  $R_d >> r_{el}$ with
a small  surface and electrostatic energy (the last term in the curly 
brackets). Using the density constraint to 
eliminate $n_{el}$ in favor of $x$ we find that for small $x$ the 
free energy behaves as   $ n^2 (-5 r_0^2 + 6 R_d^2) x$ with  $r_0$
given by
$n\equiv 3/(4 \pi r_0^3)$. Clearly to have a minimum for small $x>0$ we need
 $R_d< r_0$ and the model does not apply. The full expression for the 
free energy taking into account the electron kinetic energy gives an
even smaller sloop for the dependence of $f$ on $x$ so that an even 
smaller  drop radius is obtained. We could still have drops with a
 finite electronic densities in both the drop and the host phases. 
In this way one can reduce the
mixing energy because it depends on  the density difference $n_B-n_A$
[Eq.~(\ref{eq:em})]. One could expect to find
a solution close to the critical density for zero pressure $n_{el}^0$. 
We have searched for such a solution assuming 
$E_{el} \sim (n-n_{el}^0)^2$. It has higher energy than the uniform solution.

We can conclude that a 3D electron-jellium model  is not unstable
towards mesoscopic or macroscopic drop formation and density regions 
where the  compressibility of the electron gas is negative are physically 
accessibly. Interestingly
a negative compressibility is  actually observed for the 2D electron 
gas.\cite{eis92sha96} 
Our result stems from the fact that both the energy gain coming
from the MC and the energy cost have the same electrostatic 
origin. Of course we cannot discard instabilities that can occur at a 
microscopic scale.


\section{Frustrated phase separation in the 3D $\lowercase{t}-J$ model.}
\label{sec:stron}
In the last few years it has
become clear that many of the strongly correlated models used to describe
high temperature superconductors exhibits PS in some regions of
parameter space.\cite{mul92,sig93} Due to the strong anisotropy of 
these materials usually 2D models are consider. 
 In this section we apply the idea of a  Wigner crystal of drops 
to PS in models of strongly correlated electrons on a lattice. 
We will consider, for simplicity and homogeneity with I and the other sections 
of the paper,
isotropic 3D lattice models. We expect however that the results
will remain qualitative valid even for 2D models. Needless to say 
that the 3D
models are interesting on its own right given the large class of strongly
correlated materials where anisotropy is not important like doped C$_{60}$,
magnetoresistant manganites, etc. 

Usually in strongly correlated lattice
models the Coulomb interaction is truncated to a distance of 
a few lattice sites and often only the on-site Hubbard $U$ term is kept.
 The underlying assumption is that in a uniform ground state most
of the interesting physics is governed by the short range interactions and
that the effect of the long range interactions can be absorbed in the
 Madelung potential trough a proper Hartree renormalization of the 
on-site energies. However in a non uniform ground state  the long range 
part of the interaction has an important role even at the Hartree level. 
A simple way to take this into account is to maintain  the
  relevant 
short range interactions (e.g. the Hubbard $U$), to evaluate the energy of the 
intrinsic $A$ or $B$ phases and to add the electrostatic and surface
 contribution of  the drops to the total free energy.
 This means that we are still neglecting the Coulomb interaction at 
distances larger than the lattice constant $a$, as in the usual 
Hubbard model but we keep the Coulomb repulsion for mesoscopic distances
of the order of the inhomogeneity scale. In other words in the Fourier 
transform of the Coulomb potential, $4\pi e^2/q^2$, we maintain
 terms with  wave vector $q$ close to $q=0$  that do not cancel with the 
background and hence give a large contribution to the energy.

As an example of the relevance of this approach for strongly correlated 
systems we  focus on the  $t-J$ model, one of the more often used models in 
the cuprates. The Hamiltonian is given by:
$$
H=-t \sum_{<ij>,s} c^{\dagger}_{i,s} c^{\dagger}_{j,s} 
+ J \sum_{<ij>} ({\bf S}_i .{\bf S}_j-\frac14 n_i n_j)
$$
where $c^{\dagger}_{i,s}$ creates an electron of spin $s$ on the site $i$. 
$n_i$ and ${\bf S}_i $ being the electron number and spin operators 
respectively. Double occupation is not allowed and summations are extended   
to  nearest neighbour  of a 3D cubic lattice. 
The large $J/t$  limit has been studied in detail in 
two,\cite{eme90c,cal98,whi00} and more  dimensions based on a large $d$
expansion.\cite{car98}
We study the limiting case of small number of electrons 
(hole doping close to one). This is not particularly relevant for the
cuprates but illustrates the issue of frustrated PS in a strongly 
correlated system.

\subsection{Maxwell Construction analysis}

We start by reviewing the usual MC arguments\cite{eme90c,car98} 
in the absence of LRC  adapted to the 3D case.
The antiferromagnetic phase at half filling, hereafter the $B$ phase,
can be model by an incompressible phase with one electron per site.
the density is given by $n_{B}=n_{B}^{0}=1/a^{3}$.
Our densities refer to real electrons, not to holes. The energy is:
\begin{equation}
f_{B}=f_{B}^{0}=-3bJn_{B}^{0}  \label{eq:faf}
\end{equation}
where $bJ$ is the magnetic energy per bond. From estimates of the ground
state energy in the 3D Heisenberg model\cite{car98} one finds  $b=0.550$.

Two different situations are found for the PS.
For very large $J/t$  one finds 
 PS between the AF phase and  the electron vacuum (AF+V).
In this case we call the vacuum the $A$ phase with  $f_A=0$.
 Reducing
$J/t$ one finds PS between the AF phase and a dilute metal of 
electrons (AF+M). In this case we call the metal the $A$ phase with energy:
\begin{equation}
f_A(n_{A})=-6tn_{A}+\frac{3^{5/3}}{5}\pi^{4/3}a^{2}n_{A}^{5/3}t
\label{eq:fmetal}
\end{equation}
Here $t$ is the hopping matrix element and we have used the effective mass 
approximation in the  dispersion relation of the low density limit of the 
$t-J$ model.\cite{eme90c,car98}

\begin{figure}[tbp]
\epsfxsize=9cm
$$
\epsfbox{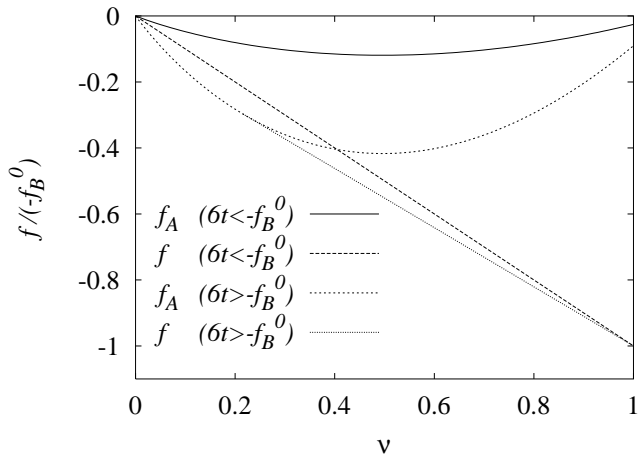}
$$
\caption{Free energy normalized to the incompressible phase free
energy without LRC and (from top to bottom close to the origin)  the metal
with  $6t=0.6|f_{B}^{0}|$, PS between the AF and vacuum (AF+V), 
the metal with $6t=2.1|f_{B}^{0}|$, MC phase separation 
between the AF and the metal (AF+M). ($n_B^0\equiv a\equiv 1$)}
\label{fig:finfmetdn}
\end{figure}

In Fig.~\ref{fig:finfmetdn} we show the free energies in the absence of LRC 
and different values of $t/(-f_{B}^{0})$. The total free energy is
given by Eq.~(\ref{eq:fdx}) with $e_m=0$. 
Since densities are assumed to be low
we can neglect the short range interaction between the electrons.
We define the number of electrons per
unit cell $\nu\equiv n/n_B^0$. In addition we set $a\equiv 1$ and restore it 
when convenient for clarity. 

For very large  $J/t$ the AF+V solution is the lowest in energy. Indeed in
Fig.~\ref{fig:finfmetdn} the upper curve corresponding to the uniform 
metallic energy does not intersects the $AF+V$ line, $f=\nu f_B^0$ 
corresponding to MC between the $\nu=0$ and $\nu=1$ points.  Decreasing the 
value of  $J/t$ when the chemical potential of the metal fulfills 
 $\mu_{A}(0)(=-6t)<f_{B}^{0}$ i.e. $t>bJ/2$ 
the metallic free energy intersects the AF+V free energy 
 at some finite density and the lowest energy state is achieved
by doing MC between the antiferromagnet and the metal. In this case as 
shown in Fig.~\ref{fig:finfmetdn},  
one finds a pure metallic phase at small density and MC phase
separation between the AF and the metal for larger density.
In Fig.~\ref{fig:tcdn} we show the phase diagram deduced form this 
analysis. The dilute metal can be unstable towards a  gas of bound 
particles.\cite{eme90c,car98} Here we do not consider this effect for 
simplicity.

In the next two subsections we analyze the effect of the LRC interaction 
on the AF+V PS and the AF+M PS. Since the electronic free energy has a simple
form we solve the equations exactly rather than  making 
a linearization as in I.

\subsection{ Drops of an incompressible phase in vacuum (AF+V)}
\label{sec:afv}
As shown above, in the absence of LRC interaction,
 this case is realized in the large $J/t$ limit.
Now we generalize the above discussion with the inclusion of LRC
and surface energy effects. 
 The $A$ phase is the electron vacuum (V)  
($n_{A}=0$ and $f_A=0$) and  the 
$B$ phase is the AF with one electron per site, $n_{B}=n_{B}^{0}=1/a^{3}$
and energy given by Eq.~(\ref{eq:faf}).
The total free energy is given by Eq.~(\ref{eq:fdx}) with the above 
replacements.
An expansion of the densities around the MC solution  
(Sec.~III of paper I) gives a trivial result since $\lambda=0$ 
(notice that $k_m=0$) 
and the densities are fixed at the MC values [Eq.~(29) of paper I]. 
However this is a peculiar limit. In fact 
 as we will show below the total free energy does not coincide 
with the MC free energy  because of the mixing energy. 
Since the densities are fixed only the radius has to be determined which is 
given  by  Eq.~(8) of paper I.

The surface energy of the AF is given by
the energy cost to cut a bond divided the associated surface $\sigma
=bJ/a^{2}$ and the volume fraction is determined by the constraint 
$x=n/n_{B}^{0}=\nu$. Inserting this in Eq.~(8) of paper I we
obtain: 
\begin{equation}
R_{d}=\left[ \frac{15}{4\pi }\frac{bJ \epsilon_0}{(2-3\nu^{1/3}+\nu) e^{2}/a}\right] ^{1/3}a
\end{equation}
 As long as the dielectric constant is
sufficiently large the radius is much larger than the lattice spacing and our
approximations are valid. 

By writing the free energy in dimensionless variables we can define a
coupling constant that will determine the transition form the AF+V solution 
and the AF+M solution in the presence of LRC. It  is given by: 
\begin{equation}
\alpha =\frac{3}{-f_{B}^{0}}\left[ \frac{9\pi e^2\sigma^2(n_{B}^{0})^2 }{5\epsilon_0}
\right]^{1/3}.  \label{eq:al}
\end{equation}

Inserting the parameters for AF drops in Eq.~(\ref{eq:al}) we find:
\begin{equation}
\alpha ^{3}=\frac{9\pi }{5}\frac{e^{2}/a}{bJ\epsilon_0}
\end{equation}
i.e. the ratio of a Coulomb energy to a magnetic energy. 

Imposing that $R_{d}>a$ for $\nu=0$ one finds 
$\alpha<3/2$ so we will 
concentrate on this range of coupling. 

\begin{figure}[tbp]
\epsfxsize=9cm
$$
\epsfbox{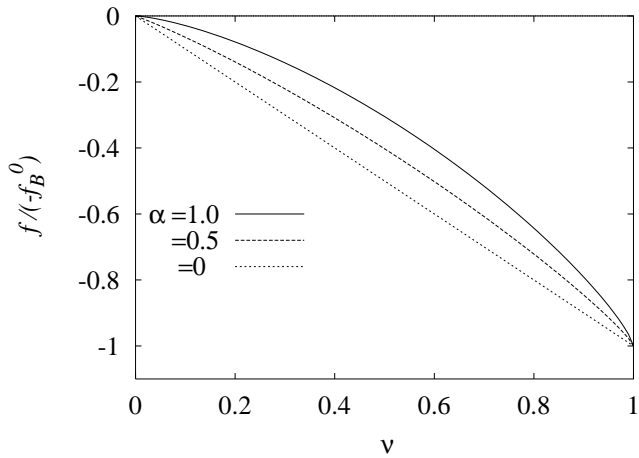}
$$
\caption{$f/(-f_{B}^{0})$ for a phase of drops of an incompressible phase in
vacuum for $\protect\alpha =0,0.5,1,2$ (from bottom to top) vs. $\nu$. }
\label{fig:findn}
\end{figure}

From Eqs.~(\ref{eq:fdx}),(\ref{eq:udx}) we obtain the free energy as a 
function of density: 
\begin{equation}
f(\nu)=[-\nu +\left(\frac{5}{243 \pi}\right)^{1/3} \alpha  u(\nu)   ](-f_{B}^{0})  \label{eq:fin}
\end{equation}
The first term in the square brackets is the bulk contribution and the
second term is the mixing energy contribution.  
This is plotted in Fig.~\ref{fig:findn} for different values of the coupling 
$\alpha $. As in the above the results are only rigorously valid
for small $x$ ( $=\nu$ ).

The effect of the  mixing energy is to bend upwards the AF+V free energy
(see Fig.~\ref{fig:findn}) so that the metallic phase can become stable with
a lower value of $t$ with respect with the case with no LRC force
 (compare with  Fig.~\ref{fig:finfmetdn}).   
One can show that for  $\alpha >1$ the AF+V solution is never stable
and one has either a uniform metal of an AF+M solution depending on doping. 
For $\alpha <1$ the drops can coexist with a metal or not  
depending on the value of $t/J$. 
We will analyze the competition with the AF+M solution in the next section.

\subsection{ Drops of an incompressible phase in a metallic host (AF+M)}

Reducing $J/t$  at some point the solution of the previous section  (AF+V) 
will not be stable any more. This has already been shown in the absence of
LRC interaction ($\alpha =0$). We consider now the $A$ phase to be the metal.

The surface energy will have now density dependent  contributions 
coming from the metal. However,
since we are in the low density limit the surface energy will be 
dominated by the magnetic surface energy described in the previous case
and can be taken as constant. The AF+V solution is then not any more stable when 
$\mu_{A}(0)=-6t<\mu _{AF+V}(0)=(\alpha -1)(-f_{B}^{0})$.

\begin{figure}[tbp]
\epsfxsize=9cm
$$
\epsfbox{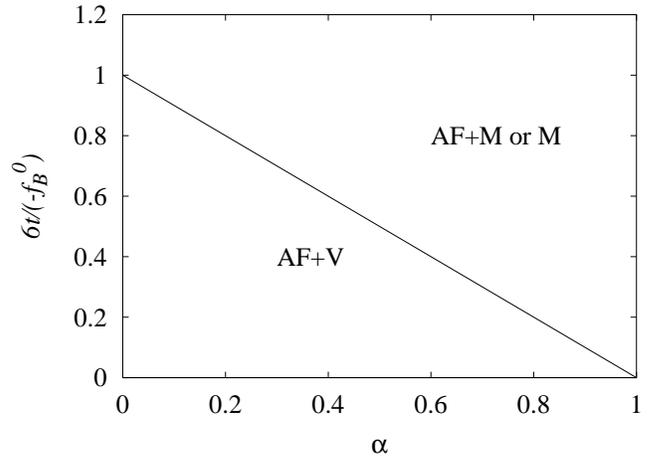}
$$
\caption{Locus of stability of the AF drops in vacuum (AF+V) in the 
$t-\alpha$ plane.
  Above the line the more stable solution depends on density.   }
\label{fig:tcdal}
\end{figure}

In Fig.~\ref{fig:tcdal} we show the locus of stability of the AF+V solution in 
the $t-\alpha $ plane. Above the line the stable solution is either a uniform 
metal or drops of AF in the metal depending on the density.
In Ref.~\onlinecite{car98} the ratio of $J/t$ below  
which  the AF+V solution is not stable for $\alpha=0$
 is called $Y_c$. In 2D they found $Y_c(0)=3.4367$  and $Y_c(0)\rightarrow 4$ 
for  $d\rightarrow \infty$.\cite{car98} Using their estimate of the 3D AF 
energy we have $Y_c(0)=3.637$. Fig.~\ref{fig:tcdal} shows
that $Y_c$ (proportional to the critical value of $-f_B^0/6t$) 
 increases with $\alpha$. Remarkably  in the presence of LRC forces a 
smaller $t$ is enough to stabilize metallic phase regions. 
In other words we can have a situation in which without LRC forces all 
the electrons are in a self bounded AF state and as the LRC
forces are switched-on some electrons ``evaporate'' to form a dilute gas 
around the AF drops.

To solve for the AF+M drop solution 
the free energy now has to be minimized with respect
to the radius and the density of the metal subject to the  constraint  
$n=xn_{B}^{0}+(1-x)n_{A}$. 
 We are implicitly assuming that the density is not low enough to form a 
Wigner crystal of electrons. One can check that for reasonable
 parameters and increasing $\alpha$ the radius of the drop becomes of
 the order of the lattice constant much before an electronic 
 Wigner crystal can form.

\begin{figure}[tbp]
\epsfxsize=9cm
$$
\epsfbox{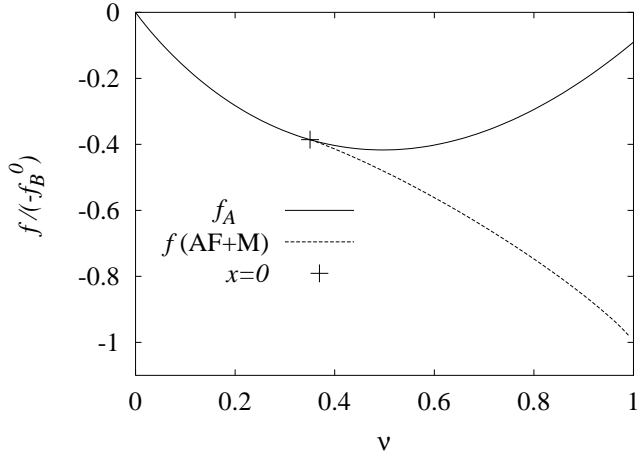}
$$
\caption{Free energy  normalized to the incompressible phase free
energy with parameters $\alpha=0.5$ and $6t=2.1|f_{B}^{0}|$.
We show the metallic free energy, the AF+V free energy 
and the AF+M free energy. 
 The cross indicates the value with $x=0$ of the AF+M drop solution.}
\label{fig:finfmetdnalp5}
\end{figure}

Above the boundary line on  Fig.~\ref{fig:tcdal}
 one finds either a uniform metal or AF+M depending on density. 
This can be seen in Fig.~\ref{fig:finfmetdnalp5} where we plot the free 
energies for $6t=2.1 |f_{B}^{0}|$ and $\alpha =0.5$. Above a certain density
$n_{bif}=\nu_{bif}/a^3$ we have coexistence of AF drops in
the metallic host. The behavior close to $n_{bif}$ is very similar
to the behavior for parabolic free energies in the UDA of paper I.

\begin{figure}[tbp]
\epsfxsize=9cm
$$
\epsfbox{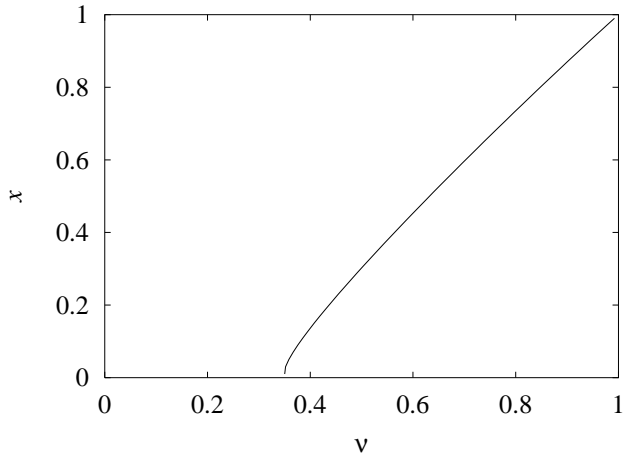} 
$$
\caption{Volume fraction vs. $\nu$ for AF drops in a metallic host and
parameters as in Fig.~\protect\ref{fig:finfmetdnalp5}.}
\label{fig:xdnalp5}
\end{figure}

Here also  there is a bifurcation of the solution and increasing the density,
the AF drops appear with a nonzero value of the volume fraction. 
 However with the present parameters 
the initial volume fraction is very small (See Fig.~\ref{fig:xdnalp5}).

\begin{figure}[tbp]
\epsfxsize=9cm
$$
\epsfbox{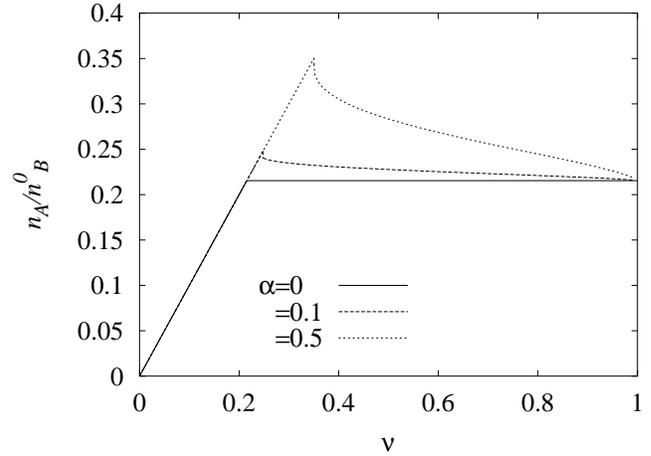} 
$$
\caption{Density in the metallic host  vs. total density $\nu$
for  $6t=2.1|f_{B}^{0}|$ and different values of $\alpha$.}
\label{fig:nsdn}
\end{figure}

As the $B$ density grows the $n_A$ density decreases due to the
 effect discussed in paper I. The $B$ density is kept constant at $n_B^0$ due
to the incompressibility. In Fig.~\ref{fig:nsdn} we show this behavior. 
In real systems this effect can be detected through physical properties
which depend on the local densities as is disused below for the manganites.

\begin{figure}[tbp]
\epsfxsize=9cm
$$
\epsfbox{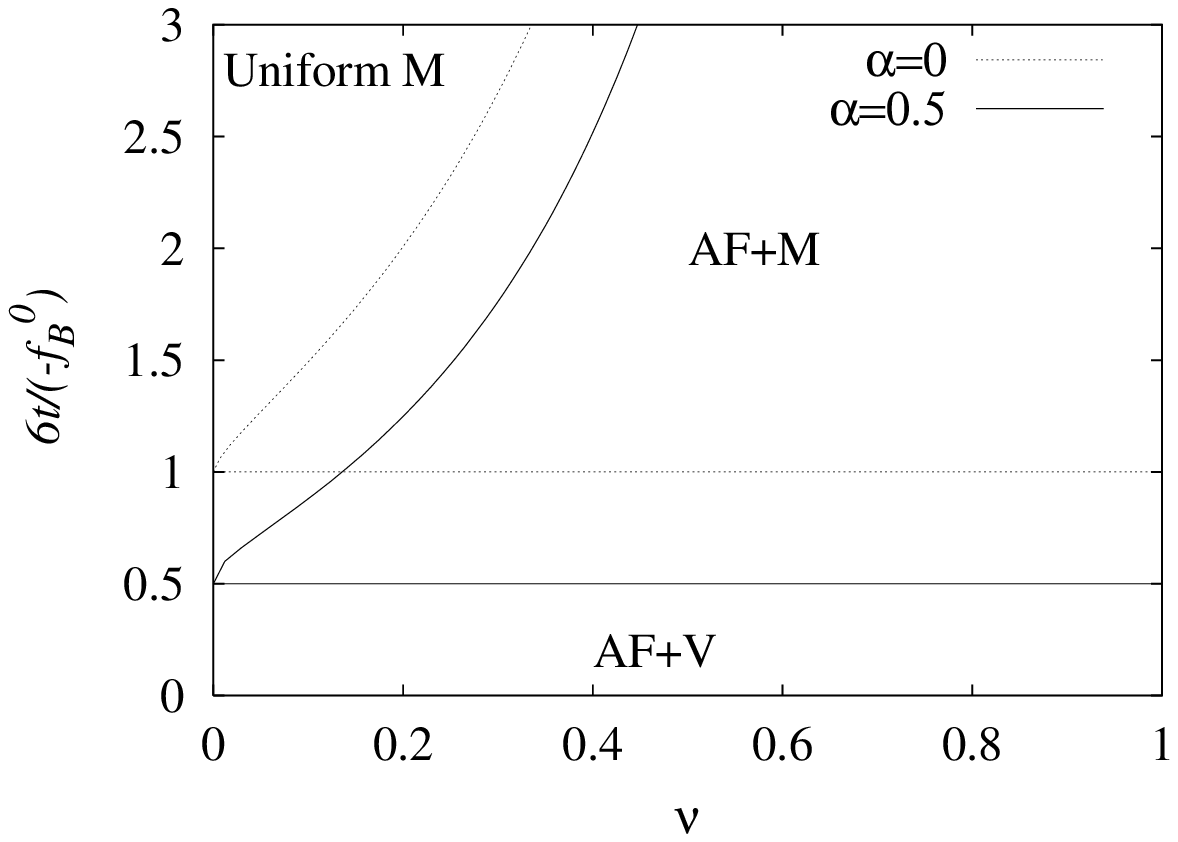} 
$$
\caption{Phase diagram for the large $J/t$ limit of the 3D $t-J$ model
without LRC ($\alpha=0$) and with a small LRC ($\alpha=0.5$). 
The high density part has to be taken with care since the 
density of electrons can be large in the metal so that short range 
interactions within the metallic phase  
cannot be neglected any more, and also drops of AF loose sense.}
\label{fig:tcdn}
\end{figure}

In Fig.~\ref{fig:tcdn} we show the phase diagram in the absence of
LRC force ($\alpha=0$) and for
$\alpha=0.5$. We see that a portion of the phase diagram in which a uniform
solution is unstable towards PS without LRC, 
for $\alpha>0$ becomes stable and the AF+M solution extends its 
region with respect to to the AF+V solution due to the ``evaporation'' 
effect.

\section{Application to the Manganites}
\label{sec:man}
As a further application we consider the   magnetoresistant 
manganites\cite{mor99} like La$_{1-y}$Ca$_y$MnO$_3$. 
In the last years strong experimental evidence has accumulated indicating 
that inhomogeneous phase separation  occurs in these  
materials in certain regions of parameter 
space.\cite{lyn96,ter97,cox98,hen98,ueh99,pap00}

At  $y=0$ all Mn have formal valence $3+$. Each ion has three electrons in 
$t_{2g}$ orbitals and one electron in an $e_g$ orbital.
The four electrons spins are  all  parallel 
due to the strong Hund's rule coupling forming an $S=2$ spin. 
The  spins of different  Mn$^{3+}$
ions form an antiferromagnetic (AF) phase due to the superexchange interaction
and the system is an insulator.   As the $e_g$ band is doped, mobile holes tend
to align the $S=3/2$ core ($t_{2g}$)   spins in different ions because 
this maximizes the transfer integral and minimize the holes kinetic 
energy leading to a ferromagnetic state. This is the so called double exchange 
(DE) mechanism.\cite{zen51,and55,gen60}

Experimentally one finds indeed the ferromagnetic state but close to 
$y=0.5$ a new  charge ordered (CO) insulating  phase  with a chessboard 
 structure  of Mn$^{3+}$ and Mn$^{4+}$  is stabilized. The CO phase is
not predicted by the conventional DE but does appear in more recent theories
incorporating Mn-Mn Coulomb repulsion\cite{kag99} or orbital 
degrees of freedom.\cite{bri99b}

Close to $y=0$ and $y=0.5$  the  metallic ferromagnetic (FM) phase 
competes with the corresponding insulating phase. The drop state due
to the competition  between the  $y=0$ AF phase and the metallic phase 
taking into account
the LRC interaction has been studied theoretically  by Nagaev and 
collaborators.\cite{nag83,nag98}  Evidence for such a phase has recently
been found in neutron scattering experiments.\cite{hen98}

Here we will analyze the competition of the CO 
phase with the FM phase close to $y=0.5$ and show that
a phase separated state can explain the puzzling 
maximum of the Curie temperature $T_c(y)$ at  $y \sim 0.35$.\cite{sch95,kim00} 
On the contrary, conventional DE would predict that the 
Curie temperature is maximum at half doping ($y=0.5$) because for this
filling the kinetic energy of the holes is maximized.

We will consider a mixed state in which the $A$ phase is the ferromagnetic metal 
(FM) and the $B$ phase is the charge order state  at $y=0.5$ which 
corresponds to inverse specific
volume   $n_B^0=0.5/a^3$. In the following the  densities refer to holes
(i.e. the concentration of Mn$^{4+}$ ions).  

In the FM phase the core spins of the Mn ions are fully polarized 
and the mobile holes have the maximum bandwidth $W$. 
In order to model the FM in a simple fashion we follow Varma\cite{var96} and
take a  flat density of states with bandwidth $W$.
The FM free energy at $T=0$ is then given by the cohesive energy of the 
holes in the fully polarized state:
 \begin{equation}
f_A(n_A)=\frac{W a^3}2 (n_A-n_B^0)^2.  
\label{eq:ffm}
\end{equation}
We have chosen to measure the single particle energies from the center of the 
band and we have dropped a constant which can be absorbed in the free energy
constant of the B phase $f_B^0$. 
At finite temperatures one has to consider the  entropy contribution to 
the free
energy. However for a given temperature one can expand the full A free energy 
around the $n_B^0$ density and an expression like Eq.~(\ref{eq:ffm}) is still
valid with an effective temperature dependent $W$.

The  CO state can be modeled as a doped incompressible phase around the  
inverse specific volume $n_B^0$. The free energy at $T=0$ can be put as:
 \begin{equation}
f_B(n_B)= \frac{E_G}2 |n_B-n_B^0|+ e_0 (n_B-n_B^0) + f_B^0.
\label{eq:fco}
\end{equation}
$f_B^0$ measures the difference in free
energy per hole 
between the CO state and the FM state at $y=0.5$ ($n=n_B^0$) and
$e_0$ controls the difference in chemical potentials of the two phases. 
$E_G$  is the gap in the charge order state. i.e. the difference
between the energies to create defects with one added hole and 
one removed hole without destroying the CO state. (It should be 
of the order of the activation energy in the transport properties of a 
pure CO state). The dip in the free energy at $n=n_B^0$ will become rounded
 with temperature. For simplicity we will neglect this effect. For 
temperatures much smaller than the gap this is a good approximation.
 Even if the temperature gets comparable to the gap a small 
rounding of the CO free
energy close to $n_B=n_B^0$ will not affect significantly the results
close to the density at which drops first appear ($n_{bif}$).  

The  chemical potential of the CO state at $T=0$ is given by:
\begin{eqnarray}
  \label{eq:muco}
  \mu^+&=&e_0+E_g/2  \;\;\;\;\;\;\;\;\;\; n_B>n_B^0 \\
  \mu^-&=&e_0-E_g/2  \;\;\;\;\;\;\;\;\;\; n_B<n_B^0 . 
\end{eqnarray}
By construction the discontinuity at $n_B^0$ is equal to the gap as it should 
be. 
$\mu^-$ is the energy to create a  Mn$^{3+}$ defect in the CO state. i.e. it 
is the energy to remove a hole in the CO state. This single particle energy is
 measured from the same reference energy as the one used for the $A$ phase in
Eq.~(\ref{eq:ffm}). This fixes the value of $e_0$. 

\begin{figure}[tbp]
\epsfxsize=9cm
$$
\epsfbox{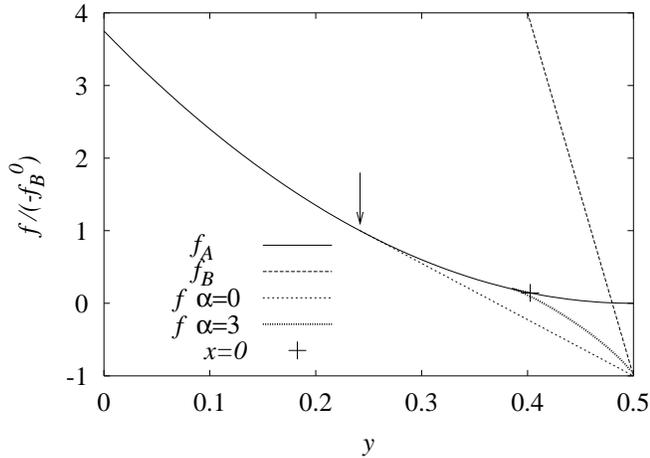}
$$
\caption{Free energy  normalized to the incompressible phase free
energy at $y=0.5$ with a  bandwidth $W=15 |f_B^0|/n_B^0$.
We show the FM free energy ($f_A$), the CO free energy with a large 
negative value of $\mu^-$,  
the FM+CO free energy for  $\alpha=0$ (MC),
and for $\alpha=3$.  The cross indicates the value with $x=0$ of
 the FM+CO drop solution. The arrow indicates the same for MC ($y_0$).}
\label{fig:fcofmdnal3wn15}
\end{figure}

In Fig.~\ref{fig:fcofmdnal3wn15} we have plotted $f_A$, $f_B$ as a function 
of $y$. We constructed the  free energies for the uniform phases  
phenomenologically, by relying on the metallic and insulating character of 
each phase and on the 
fact that due to the different magnetic symmetry they cannot be joint with
continuity but a level crossing should occur as a function of $y$. 
It is interesting to note that a recent microscopic model gives
 practically the same energy scheme as a function of doping.\cite{bri99b}

In Fig.~\ref{fig:fcofmdnal3wn15} we report also the MC and the free 
energy for the drop solution for $\alpha=3$ and
 $W'\equiv W n_B^0/ (-f_0^B)=15$ (thick line). 
 The coupling constant  $\alpha$ is defined in Eq.~(\ref{eq:al}).
$W'$ measures the effective bandwidth in units of  $(-f_0^B)$.

For the sake of simplicity we assume that the slope of $f_B$ for $n_B<n_B^0$ 
(i.e.  $\mu^-$) is so large that $f_B$ never crosses the drop solution. 
Under these simplifying condition one of the phases involved in phase 
separation is always the defect free CO state at $y=0.5$. In this situation 
$E_g$ and $e_0$ do not enter into the problem and a precise description 
of $f_B(n_B)$ is not needed. For example we have neglected 
 the kinetic energy  of the defects  which will give some curvature
 to $f_B(n_B)$ but will not change the present picture.

Alternatively to $\alpha$ we could use the coupling constant $\lambda$ defined
in Sec.~III of paper I since the FM free energy is parabolic 
($k_A^{-1}=k_m^{-1}=W a^3$) and the CO free energy can be consider as the 
$k_B\rightarrow 0$ limit of a parabola. The two coupling constants are 
related by
\begin{equation}
  \label{eq:ladeal}
  \lambda=\frac{2^{2/3}}{3}\frac{\alpha}{(W')^{1/3}}. 
\end{equation}
Specifically $\lambda=0.64$ for $\alpha=3$ and $W'=15$. Notice however that here 
(as in the previous section) we can introduce  two dimensionless parameters. One 
is $\alpha$ (or alternatively $\lambda$)  and the other
is $W'$. The latter  plays the same role as $t/J$ in the previous 
section. In particular it fixes the MC densities as follows. In the absence  
of LRC interaction ($\alpha=0$) and for 
 $W'<4$ the $f_A$ parabola is too flat and PS between the 
FM and the CO state is not possible.  One gets PS between the
 CO state and vacuum (this is similar to the AF+V PS considered in Sec.~\ref{sec:afv}). 
For  $W'> 4$ Maxwell construction gives PS between FM and CO
with the  critical doping given by:
\begin{eqnarray}
    \label{eq:y0}
    y_0= \frac12 -\frac{1}{\sqrt{W'}}    
\end{eqnarray}
In Fig.~\ref{fig:fcofmdnal3wn15} the value of $y_0$ is indicated by an arrow. 

\begin{figure}[tbp]
\epsfxsize=9cm
$$
\epsfbox{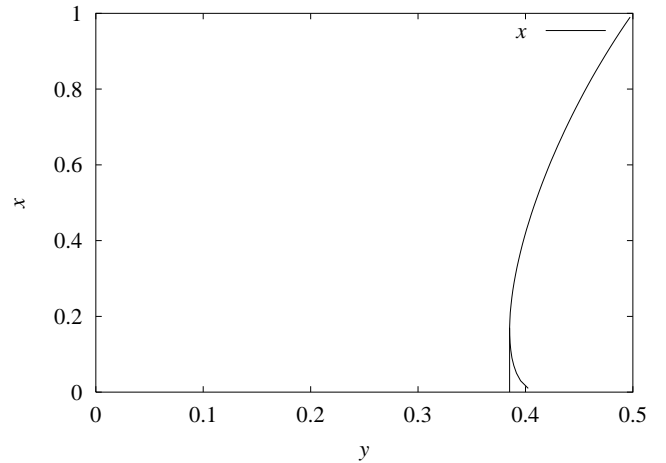}
$$
\caption{Volume fraction vs. $y$ for CO drops in a FM host and 
parameters as in Fig.~\protect\ref{fig:fcofmdnal3wn15}. The lower branch
 close to $y\sim 0.4$ is unphysical.}
\label{fig:xdnal3wn15}
\end{figure}

In the presence of LRC interaction the range of coexistence contracts with respect
to the Maxwell construction case. The transition from the FM to the
drop solution is quite abrupt at $y_{bif}=0.38$ (for $\alpha=3$)
with a substantial jump of the volume fraction from zero to a finite value $x_{bif}=0.17$.
(See Fig.~\ref{fig:xdnal3wn15}).

\begin{figure}[tbp]
\epsfxsize=9cm
$$
\epsfbox{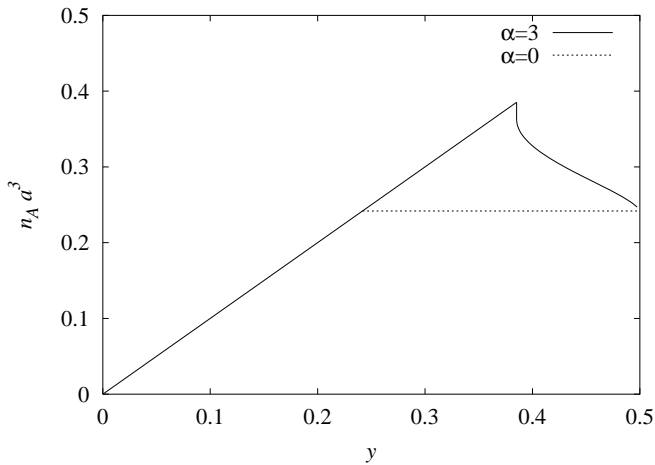}
$$
\caption{Density in the metallic host  vs. doping  $y$ for
parameters as in Fig.~\protect\ref{fig:fcofmdnal3wn15}.}
\label{fig:nsdnal3wn15}
\end{figure}

In Fig.~\ref{fig:nsdnal3wn15} we show the local density inside the
metallic region. For $y<y_{bif}$ the stable phase is uniform FM 
and the total density coincides with the nominal density
$n=y/a^3$. For $y>y_{bif}$ the drop solution is stable and the
density in the metallic region decreases with increasing nominal density.
As discussed in paper I,
 in deriving this effect it is important 
that the density dependence of the surface energy can be neglected. 
The strongest dependence of the surface energy is expected to arise from the kinetic energy 
of the metal. However, this dependence is important close to $y\sim 1$ and  $y\sim 0$
and can be safely  neglected close to  $y\sim 0.5$.

The decrease of local density with increase of global density 
 can explain the  non-monotonous dependence 
of Curie temperature $T_c$ on $y$. Since the ferromagnetic  interaction between the core 
Mn spins is mediated by the conduction electrons through the double 
exchange mechanism one expects the Curie temperature to be a monotonous
increasing function of  the local metallic density $n_A(<n_B^0)$  of the FM phase.
We associate the region in which the Curie temperature increases with doping,
i.e. the ``normal'' region (roughly $0.1<y<0.33$ for  La$_{1-y}$Ca$_y$MnO$_3$),
 with a uniform FM phase 
and the ``anomalous'' regions in which the Curie temperature decreases with 
doping with a drop state. In the latter state $n_A$ decreases with doping 
and this gives the anomalous behavior of $T_c$ as a function of doping.

To be more specific
 we assume the following simple form for the dependence of the Curie
temperature on the local FM density:
\begin{eqnarray}
  \label{eq:tcdn}
 \frac{ t_c(a^3 n_A)-t_c(0)}{t_c(0.5)-t_c(0)}=4  (1-a^3 n_A)*a^3 n_A 
\end{eqnarray}
We are using uncapitalized $t$ for the local Curie temperature of the 
FM phase  to distinguish it from the true Curie temperature
of the system which is a function of  the overall doping $T_c(y)$.   
A similar form as  Eq.~(\ref{eq:tcdn}) with $t_c(0)=0$ was derived by Varma 
for a uniform FM phase.\cite{var96} More sophisticate
treatments also give a form roughly parabolic with $t_c(0)>0$.\cite{oka00}

 For a uniform FM phase $y=n_A a^3$, $T_c(y)=t_c(y)$.
This fits correctly the experimental data in the   normal region. 
We can use this fit to fix the parameters in 
Eq.~(\ref{eq:tcdn}). For  La$_{1-y}$Ca$_y$MnO$_3$ one obtains 
$t_c(0)\sim 80K $ and $t_c(0.5)\sim 300K$. Close to $y=0.5$ 
we have the anomalous 
behavior and the measured $T_c$ differs considerably from
 $t_c(0.5)$, the Curie temperature of an hypothetically uniform 
phase. For example, experimentally $T_c(0.5)\sim 225 K$.

\begin{figure}[tbp]
\epsfxsize=9cm
$$
\epsfbox{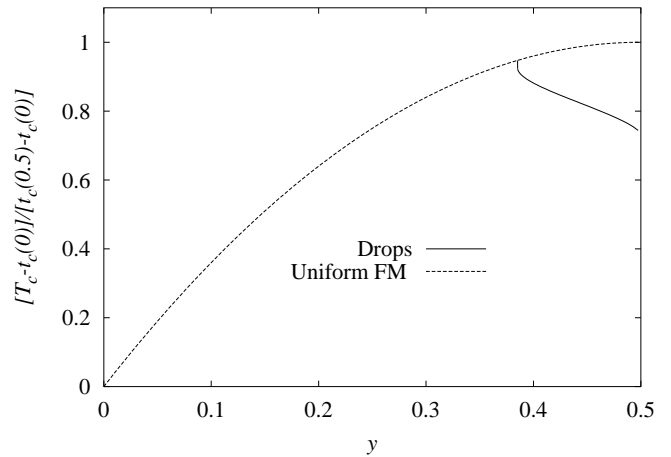}
$$
\caption{$T_c$ of the FM host minus $t_c(0)$ normalized to $t_c(0.5)-t_c(0)$
as a function of doping  $y$.  
Parameters are  as in Fig.~\protect\ref{fig:fcofmdnal3wn15}.
We show the $T_c$ in the uniform solution and in the drop solution. 
}
\label{fig:tcdnal3wn15}
\end{figure}

From the known $n_A(y)$ (Fig.~\ref{fig:nsdnal3wn15}) and Eq.~(\ref{eq:tcdn})
we compute $T_c(y)\equiv t_c[a^3 n_A(y)]$. This curve (which is quite similar 
to the experimental one) is shown in Fig.~\ref{fig:tcdnal3wn15}. 
Indeed we  see that the drop solution combined with the uniform solution
for $y<y_{bif}$ gives a non-monotonous behavior of  $T_c(y)$.
 In evaluating the theoretical curve to be 
compared with the experimental data we fix the values of $\alpha$ and
$W'$ in the
following way. We  associate the 
experimental maximum in $T_c$ with the bifurcation point. i.e. the doping
at which the  uniform solution  switches  to the drop solution. This gives 
us an experimental value of the bifurcation doping $y_{bif}\sim 0.38$. 
From the experimental data we also obtain the depression of the 
Curie temperature $[T_c(0.5)-t_c(0)]/[t_c(0.5)-t_c(0)]\sim 0.66$.
With these two dimensionless numbers we obtain the dimensionless 
parameters of our theory and find   $\alpha=3$ and $W'=15$ i.e. the
values that we have been using in the present section. 
 A rough  microscopic estimate of these  parameters is given in 
Appendix~\ref{app:param} to show that indeed the above values are 
reasonable for the manganites. 

A similar 
behavior as the one discussed here for $y\sim 0.5$ is observed  close 
to  $y\sim 0$. We speculate that this is due to the same  general
phenomena involving inhomogeneous phase separation between the
insulator at $y=0$ and the ferromagnetic metal. 

\section{Conclusions}
\label{sec:con}

In this work we have applied the ideas developed in I to three different 
physical systems.  First we analyzed well known apparent instabilities 
in the low density limit of the jellium model. Usual instability
criteria like a negative compressibility are formulated for a neutral
system and should be taken with care in a charged system with a
compensating background. In a charged system 
an instability of the kind implied by a MC analysis is a
necessary but not sufficient  condition for mesoscopic phase
separation. In fact we have seen that for the Wigner crystal of electrons
the Coulomb strength and the 
surface energy balance in such a way that large drops are not
possible. This is basically due to the fact that 
the energy gain from MC and the  energy cost due to the LRC interaction 
and surface energy have all the same electrostatic origin. This prevent 
the mixing energy to be small compared to the MC energy as required in I
to have mesoscopic PS. We mention that in the 2D electron gas (not consider 
here) there is evidence both for a density region of negative compresibilities 
and at lower densities phase separation.\cite{ila00,eis92sha96} For the 
latter however the effect of disorder not consider by us may be crucial.

The study of $t-J$ model illustrates the importance of the mixing energy 
on determining the phase diagram. The LRC interaction tends to stabilize 
the non-separated uniform phases in the presence of a rigid
background.
Apart from this intuitive effect the LRC interaction can
also favor one mixed state over another. In fact we showed that
 the LRC interaction can transform the clusters of AF electrons 
in vacuum  into clusters of AF electrons in 
equilibrium with 
its vapor (a dilute electron gas), a phenomenon which we referred to as 
``evaporation''.

In paper I of this series we showed that the local densities of 
the two phases tend to have an anomalous behavior in the mixed state. 
Both of them tend to decrease when the global density increases. 
This non-linear effect  can affect properties of the
system which are sensitive to the local density, as we have
illustrated for the Curie temperature in the manganites. We have thus
provided an explanation to anomalies that occur  in the phase 
diagram. i.e. A decreasing Curie temperature when the CO state is
approached by varying the doping.   

\appendix

\section{Microscopic estimate of parameters in the Manganites }
\label{app:param}
In Sec.~\ref{sec:man} we   find that the parameters  $\lambda=0.64$ and $W'=15$
give a curve $T_c(y)$ similar to the experimental one. 
To decide if these parameters are reasonable one needs a microscopic 
computation. 

To evaluate $W'$  which appears in Eq.~(\ref{eq:y0}) we refer to
a recent zero temperature microscopic computation which takes
into account double exchange and orbital ordering.\cite{bri99b}
Their  Fig.~2 showing the free energy (without LRC) is quite similar
 to our $\alpha=0$ curves in Fig.~\ref{fig:fcofmdnal3wn15}.
 From there we take $y_0\sim 0.24$ which determines 
$W'\sim 15$ [Eq.~(\ref{eq:y0})] in agreement with the value
we used to fit $T_c$. 
$\lambda$ is more difficult to obtain because it requires a  microscopic 
computation of surface energies and screening effects. We parameterize
the surface energy by a dimensionless quantity $\gamma$ defined by
 $\sigma\equiv\gamma W/a^2$.
Putting $\delta_0=(0.5-y_0)/a^3$ and $k_A^{-1}=k_m^{-1}=W a^3$ in 
Eq.~(25) of paper I we get:
\begin{equation}
  \label{eq:ladal}
  \lambda=2\left(\frac{9\pi}5\right)^{1/3}\frac{\gamma^{2/3}}{(0.5-y_0)^{4/3}} 
  \left(\frac{e^2 }{\epsilon_0 a W} \right)^{1/3} 
\end{equation}
For the bandwidth we can take an estimate based on
Mattheiss's local density approximation  
(See Ref.~\onlinecite{var96}) $W=2.5eV$. 
For a cubic array of Mn with a Mn-Mn distance of 4\AA 
we get  $e^2/a=3.4eV$ for the bare Coulomb strength.
Inserting the  numerical values in the above equation we have 
$$
\lambda\sim 21\left( \frac{\gamma^2}{\epsilon_0}\right)^{1/3}
$$
One obtains $\lambda\sim 0.64$, the value we have used in Sec.~\ref{sec:man}, 
by taking $\epsilon_0\sim 100$ and $\gamma\sim 0.05$.
These are reasonable parameters considering that $\epsilon_0$
 should be understood as a static dielectric constant 
taking into account conventional dielectric screening plus 
Thomas-Fermi screening effects (Sec.~IV of paper I) and 
$\gamma W$ should be a small fraction of the bandwidth. 

We mention that since a real background is never perfectly rigid, 
a volume relaxation will also occur  inside  the drop phase. 
In general the positive
background will contract in the electron rich phase and expand in the 
electron poor phase to reduce the mismatch between the ionic positive density
and the electronic density. This is in agreement
whit the situation in Pr$_{0.7}$Ca$_{0.3}$MnO$_3$ where the electron poor 
CO phase has a larger volume than the electron rich FM phase.\cite{cox98} 
Clearly this effect has to be included in the effective definition 
 of  $\epsilon_0$.

The drop radius reads:
\begin{equation}
  \label{eq:rda}
  R_d=
\frac{3 2^{1/3} \gamma a}{\lambda[(\delta a^3)^{2} (0.5-y_0)^{4} (2-3x^{1/3}+x)]^{1/3} } 
\end{equation}
Using the above parameters we can estimate the radius at the onset 
($x_{bif}=0.17$) to be of the order of
$$
R_d\sim 10 a
$$
Correspondingly the cell radius is  $R_c=R_d/x^{1/3} \sim 18 a$. 
We see  that these scales are much larger than
the lattice constant and our approximations apply.


\begin{references}

\bibitem{mul92}
 in {\em Phase separation in cuprate superconductors}, edited by K.~A. Muller
  and G. Benedek (World Scientific, Singapore, 1992).

\bibitem{sig93}
 in {\em Phase separation in cuprate superconductors}, edited by E. Sigmund and
  K.~A. Muller (Springer-Verlag, Berlin, 1993).

\bibitem{low94}
U. L\"ow, V.~J. Emery, K. Fabricius, and S.~A. Kivelson, Phys.\ Rev.\ Lett.
  {\bf 72},  1918  (1994).

\bibitem{eme90c}
V.~J. Emery, S.~A. Kivelson, and H.~Q. Lin, Phys.\ Rev.\ Lett. {\bf 64},  475
  (1990).

\bibitem{cas95b}
C. Castellani, C. {Di Castro}, and M. Grilli, Phys.\ Rev.\ Lett. {\bf 75},
  4650  (1995).

\bibitem{nag83}
E. Nagaev, {\em {Physics of magnetic semiconductors }} (MIR, Moscow, 1983).

\bibitem{mor99}
A. Moreo, S. Yunoki, and E. Dagotto, Science {\bf 283},  2034  (1999).

\bibitem{ila00} S. Ilani {\it et al.} Phys. Rev. Lett. {\bf 84} 3133 (2000). 

\bibitem{nag98}
E. Nagaev, A.~I. Podel'shchikov, and V.~E. Zil'bewarg, J. Phys.: Condens.
  Matter {\bf 10},  9823  (1998).

\bibitem{mah90}
G.~D. Mahan, {\em Many Particle Physics} (Plenum, New York, 1990).

\bibitem{eis92sha96} J. P. Eisenstein, L. N. Pfeiffer and K. W. West, 
 Phys.\ Rev.\ Lett. {\bf 68},  674
  (1992); S. Shapira {\it et al.},  Phys.\ Rev.\ Lett. {\bf 77},  3181 (1996).

\bibitem{cal98}
M. Calandra, F. Becca, and S. Sorella, Phys.\ Rev.\ Lett. {\bf 81},  5185
  (1998).

\bibitem{whi00}
S.~R. White, Phys.\ Rev.\ B {\bf 61},  6320  (2000).

\bibitem{car98}
E. Carlson, S.~A. Kivelson, Z. Nussinov, and V. Emery, Phys.\ Rev.\ B {\bf 57},
   14704  (1998).

\bibitem{lyn96}
J.~W. Lynn {\it et~al.}, Phys.\ Rev.\ Lett. {\bf 76},  4046  (1996).

\bibitem{ter97}
J. {De Teresa {\it et al.}}, Nature (London) {\bf 386},  256  (1997).

\bibitem{cox98}
D.~E. {Cox {\it et al.}}, Phys.\ Rev.\ B {\bf 57},  3305  (1998).

\bibitem{hen98}
M. {Hennion {\it et al.}}, Phys.\ Rev.\ Lett. {\bf 81},  1957  (1998).

\bibitem{ueh99}
M. Uehara, S. Mori, C. Chen, and {S.-W. Cheong}, Nature (London) {\bf 399},
  560  (1999).

\bibitem{pap00}
G. {Papavassiliou {\it et al.}}, Phys.\ Rev.\ Lett. {\bf 84},  761  (2000).

\bibitem{zen51}
C. Zener, Phys.\ Rev. {\bf 82},  403  (1951).

\bibitem{and55}
P.~W. Anderson and H. Hasegawa, Phys.\ Rev. {\bf 100},  675  (1955).

\bibitem{gen60}
P.~G. {de Gennes}, Phys.\ Rev. {\bf 118},  141  (1960).

\bibitem{kag99}
M.~Y. Kagan, D.~I. Khomskii, and M.~V. Mostovoy, Eur. Phys. J. B {\bf 12},  217
   (1999).

\bibitem{bri99b}
J. {van der Brink}, G. Khaliullin, and D. Khomskii, Phys.\ Rev.\ Lett. {\bf
  24},  5118  (1999).

\bibitem{sch95}
P. Schiffer, A.~P. Ramirez, W. Bao, and S.-W. Cheong, Phys.\ Rev.\ Lett. {\bf
  75},  336  (1995).

\bibitem{kim00}
K.~H. kim, M. Uehara, and S.-W. Cheong, cond-mat/0004467 (unpublished).

\bibitem{var96}
C.~M. Varma, Phys.\ Rev.\ B {\bf 54},  7328  (1996).

\bibitem{oka00}
S. Okamoto, S. Ishihara, and S. Maekawa, Phys.\ Rev.\ B {\bf 61},  451  (2000).

\end{references}

\end{document}